\documentclass[aip,jcp,amsmath,amssymb,reprint,superscriptaddress,groupedaddress]{revtex4-1}

\usepackage{graphicx}
\usepackage{subfig}

\usepackage{amssymb,amsmath,amsthm,mathrsfs,comment,afterpage,accents}

\makeatletter
\def\@dotsep{4.5}
\makeatother

\usepackage{dcolumn}
\usepackage{bm}

\newcommand{\bra}[1]{\langle #1|}
\newcommand{\ket}[1]{|#1\rangle}

\begin{document}

\title{The remarkable accuracy of an $O(N^6)$ perturbative correction to opposite-spin CCSD: are triples necessary for chemical accuracy in coupled cluster?}

\author{David W. Small}

\affiliation{Molecular Graphics and Computation Facility, College of Chemistry, University of California, Berkeley, California 94720}

\date{\today}

\begin{abstract}
The focus of this work is OS-CCSD-SPT(2), which is a second-order similarity transformed perturbation theory correction to opposite spin coupled cluster singles doubles, 
where in the latter the same-spin amplitudes are removed and the opposite-spin ones are solved self consistently.
We demonstrate that, for non-multireference molecules, OS-CCSD-SPT(2) produces relative energies that rival the accuracy
of higher-order methods like CCSD(T).  For example, using PBE0 orbitals in the reference, OS-CCSD-SPT(2) exhibits a mean absolute deviation (MAD) of 0.66 kcal/mol 
with respect to CCSD(2) benchmark values for the non-multireference subset of W4-08 atomization energies (c.f.\ a MAD $>$ 6 kcal/mol for CCSD).
OS-CCSD-SPT(2) is free of empirical parameters, has an instrinsic scaling of $O(N^6)$, and makes no use of triples.
It is also naturally amenable to higher order corrections: the associated third-order correction, OS-CCSD-SPT(3), which does involve connected
triples and quadruples, exhibits a MAD of 0.44 kcal/mol for the same benchmark.
\end{abstract}

\maketitle

For its consistently high accuracy, Coupled Cluster Singles Doubles with Perturbative Triples (CCSD(T))\cite{RaghavachariCPL1989} is known as electronic structure theory's 
``gold standard''.\cite{Cramer_Essentials,RamabhadranJCTC2013,RezacJCTC2013}
Nevertheless, in its unabridged form, its range of application is severely limited by a $O(N^7)$ scaling with system size.  Over the years, the push to reduce
this burden has been quite successful, culminating in linear scaling approximations such as
DLPNO-CCSD(T),\cite{RiplingerJCP2013,GuoJCP2018,LiakosJPCA2020}
LNO-CCSD(T),\cite{RolikJCP2013,NagyJCTC2019}
CIM-CCSD(T),\cite{LiJCP2009}
DC-CCSD(T),\cite{KobayashiJCP2009}
OSV-L(T),\cite{SchutzJCP2013}
PNO-CCSD(F12*)(T),\cite{SchmitzJCTC2017}
and DEC-CCSD(T).\cite{EriksenJCTC2015}
Overall, this research has reached a mature state and these methods are now being used routinely in chemical applications.

The triples contribution is overwhelmingly rate limiting for unabridged CCSD(T), and it remains the bottleneck for the reduced-scaling approaches (see e.g.\ ref.\ 
\onlinecite{GuoJCP2018}).  The purpose of the present Letter is to investigate the possibility of using a simple new approach for bypassing the triples term altogether 
(and, in fact, reasonably approximate the effects of quadruples, too).  The following observations motivate the new approach.

Spin-component scaling (SCS), in which different overall weights are applied to opposite-spin (OS) components and same-spin (SS)
components, has been exploited to much success.\cite{GrimmeJCP2003,SzabadosJCP2006,FinkJCP2010,GrimmeWCMS2012}  
SCS generally improves the accuracy of a given approximation without increasing its computational complexity. 
These advances have made it clear that OS and SS amplitudes have important physical distinctions, and therefore they can and should be treated differently from each
other.

In another line of research, it is known that a related partition of the amplitudes has significant implications for understanding the behavior
of CCSD in strongly correlated systems.  In closed shell CCSD, the doubles substitution operator $T_2$ may be divided into 2 
well-defined doubles operators, $^SS$ and $^AS$.  These operators remove singlet and triplet pairs from the reference wave function, respectively, and replace them
with unoccupied-orbital singlet and triplet pairs, respectively.  The SS doubles are fully contained within $^AS$.
It has recently been shown that the $^AS^2$ term is the dominant source of inaccuracy for CCSD in strong correlation situations.\cite{BulikJCTC2015,LeeJCTC2017}

With the above considerations in mind, we propose the following approach, which to our knowledge has not been explored before.
  Its basis is opposite-spin CCSD (OS-CCSD).  Here, the SS amplitudes are set to 0, and the 
projection equations for single substitutions and OS doubles substitutions are solved.  Note that this differs from SCS-CCSD,\cite{TakataniJCP2008,PitonakPCCP2010}
in which the CCSD amplitudes are solved in the usual way and then the spin components are re-scaled.  OS-CCSD by itself is clearly incomplete, and here this will
be remedied by similarity transformed perturbation theory (SPT).\cite{KucharskiJCP1998}  The energies through 2nd and 3rd order may be written as

\begin{equation}
E^{(2)} = \bra{\Psi^{(1)}} \bar{H} \ket{0} = \bra{0} \bar{H} \ket{\Psi^{(1)}}
\end{equation}
and
\begin{equation}\label{E3def}
E^{(3)} = E^{(1)} + \bra{L} (\bar{H}-E^{(1)}) \ket{\Psi^{(1)}},
\end{equation}
where $\ket{0}$ is the reference determinant, $\bar{H} = e^{-T} H e^T$, 
the energy through first order is $E^{(1)} = \bra{0} \bar{H} \ket{0}$, and the (right) wavefunction expanded to first order
is 
\begin{equation}
\ket{\Psi^{(1)}} = \ket{0} + (E^{(0)}-H_0)^{-1} V \ket{0},
\end{equation}
where $H_0$ is the zeroth-order Hamiltonian, $E^{(0)}$ is the zeroth-order energy, and $V = \bar{H}-H_0$.
$\bra{L}$ is a reference + singles + doubles wave function that will be further specified below.
Because $\bar{H}$ is not Hermitian, the left first-order wavefunction, $\bra{\Psi^{(1)}} = \bra{0} + \bra{0} V (E^{(0)}-H_0)^{-1}$, is distinct from the right one.

For the OS-CCSD case, we will use the regular Fock operator $F$ for $H_0$.  More precisely, we use the occupied-occupied and virtual-virtual blocks of $F$, which enables
the use of non Hartree Fock (HF) orbitals.  The above orders of SPT applied to OS-CCSD will be denoted in this Letter
by OS-CCSD-SPT($j$), where $j$ is the SPT order (2 or 3).  In this work, for SPT(3), $\bra{L}$ is equal to $\bra{\Psi^{(1)}}$.

For the CCSD family, e.g.\ CCSD, CCD, OS-CCSD, etc., the SPT correction of $E^{(2)}$ involves a simple contraction of the singles and doubles amplitudes of $\bra{\Psi^{(1)}}$
with the full CCSD residual.  The time required for this is negligible compared to that for the formation of the residual.  For OS-CCSD-SPT(2), this contraction reduces to
going over the SS doubles substitutions, and the SS doubles amplitudes of $\bra{\Psi^{(1)}}$ are the same as the MP2 SS doubles.

$E^{(3)}$ involves additional singles and doubles
terms along with approximate connected triples and quadruples terms.  The triples term is essentially analogous to that for CCSD(T), except mainly that 
perturbative amplitudes or $\Lambda$ amplitudes (see below) are used instead of
$T$ in $\bra{L}$.  When $\bra{L} = \bra{\Psi^{(1)}}$, the quadruples term can exactly be factorized and computed in $O(N^6)$ time.\cite{KucharskiJCP1998}

The $E^{(3)}$ components are analogous to those of CCSD(2).  CCSD(2)\cite{GwaltneyCPL2000,GwaltneyJCP2001} is an
accurate SPT variant that uses $\bar{F}$, the one-electron part of $\bar{H}$,\cite{GwaltneyJCP2001}
for $H_0$.  Although CCSD(2) was originally formulated as a 2nd-order SPT, its energy may be defined using eqn.\ \ref{E3def}:
the CCSD $\Lambda$ amplitudes are used for $\bra{L}$, while a quadruples factorization is again used although 
now this is approximate.  Compared to CCSD(T), CCSD(2) is advantageous in that it uses proper left-eigenvector amplitudes for 
$\bra{L}$,\cite{CrawfordIJQC1998,HirataJCP2001,PiecuchJCP2005,TaubeJCP2008} 
and it includes connected quadruples.  Accordingly, in this letter, CCSD(2) will be used to benchmark the new approach.

The primary benchmark data used in this work are (electronic) atomization energies (AE),
which in general serve as a challenging benchmark test.\cite{KartonJCC2017}
For this, we employ 3 datasets: (1) the non-multireference (non-MR) subset of W4-08,\cite{KartonJPCA2008} 
which will be denoted by W4-08woMR here, (2) the diatomic 3$d$ transition metal oxides, and (3) the diatomic 3$d$ transition metal sulfides.  The latter two are subsets of
the TMD60 set,\cite{ChanJCTC2019} which originates from the work of Jensen et.\ al.\cite{JensenJCP2007}, and we obtained the geometries from ref.\ \onlinecite{HaitJCTC2019}.
These two sets will be denoted by TMD60(O) and TMD60(S) in this work.  The geometries for W4-08 were obtained from the GMTKN30\cite{GoerigkJCTC2011} database.\cite{GMTKN30website}
In addition, we will examine two simple non-covalent potential energy surfaces (PESs): the sandwich benzene dimer and Ar$_2$.

All calculations below used a development version of Q-Chem.\cite{ShaoMP2015}  All W4-08, TMD60(O), TMD60(S), and Ar$_2$ calculations used 
def2-TZVPPD\cite{RappoportJCP2010} for the basis set, 
and the benzene-dimer calculations used def2-SV(P).\cite{EichkornTCA1997}  All W4-08 and TMD60 calculations used unrestricted orbitals and were checked with 
stability analysis.\cite{SeegerJCP1977}  We will use the following statistics for benchmarking: mean deviation (MD), mean absolute deviation (MAD), 
root mean square deviation (RMSD), minimum deviation (minD), and maximum deviation (maxD).

Our first test is to determine if it is possible to replace $\bar{F}$ with its symmetric part, i.e.\ $\bar{F}_s = \frac{1}{2} (\bar{F} + \bar{F}^t)$, with superscript $t$ denoting 
the transpose.  The reason for this is that Q-Chem's CCSD(2) algorithm diagonalizes $\bar{F}$ in order to invert it, but is not compatible with complex eigenvalues, which are not 
uncommon for $\bar{F}$.  There are 111 atoms and molecules in W4-08, and excluding the one structure that has complex eigenvalues, the MAD in \emph{total}
energy between using
$\bar{F}$ and $\bar{F}_s$ in CCSD(2) (with HF orbitals) is $5.8 \times 10^{-7}$ a.u.\ and the maximum absolute deviation is $9.2 \times 10^{-6}$ a.u.  We conclude that $\bar{F}$
may readily be replaced by $\bar{F}_s$ for $H_0$ in CCSD(2).

Next, we test OS-CCSD-SPT(2).  Overall statistics for the AE data for W4-08woMR are given in Table \ref{tableW408}.

\begin{table}[t]
\caption{Error statistics for electronic atomization-energy data for the W4-08woMR set.  All results are in kcal/mol and deviation data are relative to CCSD(2) 
with the same orbitals.  CCSD(2) used $\bar{F}_s$ for $H_0$.}
\centering
\begin{tabular}{c c c c c c}
\multicolumn{6}{c}{Using HF orbitals}\\
\hline\hline
\               & \ \ MD\ \ & \ MAD\ & \ RMSD & \ minD  & \ maxD \\ [0.5ex]
\hline
OS-CCSD-SPT(2)  &  -0.81    &  0.98  &  1.34  &  -5.15  &  1.17  \\
OS-CCSD-SPT(3)  &  -0.23    &  0.49  &  0.77  &  -3.16  &  2.21  \\
CCSD            &  -6.16    &  6.16  &  7.53  & -18.19  &  0     \\
CCSD(T)         &   0.45    &  0.45  &  0.67  &  -0.17  &  3.06  \\
[1ex]
\hline\noalign{\medskip}
\multicolumn{6}{c}{Using PBE0 orbitals}\\
\hline\hline
\               & \ \ MD\ \ & \ MAD\ & \ RMSD & \ minD  & \ maxD \\ [0.5ex]
\hline
OS-CCSD-SPT(2)  &  -0.48    &  0.66  &  0.86  &  -2.79  &  1.31   \\
OS-CCSD-SPT(3)  &   0.19    &  0.44  &  0.69  &  -2.05  &  2.04   \\
CCSD            &  -6.68    &  6.68  &  8.16  & -20.00  &  0      \\
CCSD(T)         &   0.73    &  0.73  &  1.02  &  0      &  3.62   \\
[1ex]
\hline
\end{tabular}
\label{tableW408}
\end{table}

In the top half of Table \ref{tableW408}, it is evident that the average errors of HF-based OS-CCSD-SPT(2) are much lower than for CCSD.  In fact, they are 
reasonably close to the average deviations of OS-CCSD-SPT(3) and CCSD(T), which are similar to each other.
Thus, although OS-CCSD-SPT(2) improves upon CCSD, it evidently does not over-correct, in the sense that applying the next order of 
SPT, OS-CCSD-SPT(3), only increases the accuracy.

  Although W4-08woMR is nominally non-MR, there remain a few 
molecules in this subset that exhibit symmetry breaking (SB) at the HF level, which is either artifactual or indicates (at least borderline) MR 
character; either way this is problematic for perturbative corrections.  For example, for HF, the singlet molecules S$_2$O and P$_2$ have $\langle S^2 \rangle$ 
values of 0.61 and 0.73, respectively, while the doublet molecules C$_2$H and CN show values of 1.15 and 1.16, respectively.  
DFT orbitals are less susceptible to symmetry contamination than HF orbitals, and using the former for CC calculations has been shown to be an effective
way of dealing with this problem.\cite{BeranPCCP2003,VasiliuJPCA2015,FangJCTC2016}
  Using PBE0\cite{AdamoJCP1999} orbitals, S$_2$O and P$_2$ both have $\langle S^2 \rangle = 0$, while the values for C$_2$H and CN
reduce to 0.79 and 0.76, respectively.  Over the whole W4-08woMR set, the HF-based CCSD(2) and PBE0-based CCSD(2) AEs are generally quite close, 
with MAD = 0.15 kcal/mol, but there are exceptions: e.g.\ the maximum absolute deviation is 1.40 kcal/mol, which again suggests there exist some
SB or MR-related problem cases.

As a result of these considerations, we recomputed all W4-08woMR energies for the relevant CC approximations using PBE0 orbitals, and the results 
are given in the bottom half of Table \ref{tableW408}.  The deviations for PBE0-orbital OS-CCSD-SPT(2) are significantly lower than for
HF-orbital OS-CCSD-SPT(2), and with a MAD of 0.66 kcal/mol, are nearing the (chemical) accuracy level of OS-CCSD-SPT(3).  Here, CCSD(T) actually
has a slightly larger MAD of 0.73 kcal/mol.

The TMD60 structures generally exhibit significantly more spin contamination than for W4-08woMR, e.g.\ HF orbitals give
$\langle S^2 \rangle = 1.01$ for ZnO (singlet) and $\langle S^2 \rangle = 3.02$ for NiS (triplet).
We thus only report PBE0-orbital-based results here.  These are shown in Table \ref{tableTMD60}.

\begin{table}[t]
\caption{Error statistics for electronic AE data for the TMD60(O) and TMD60(S) sets.  All calculations used PBE0 orbitals, and all results are in kcal/mol.
Deviation data are relative to CCSD(2) (which used $\bar{F}_s$ for $H_0$).}
\centering
\begin{tabular}{c c c c c c}
\multicolumn{6}{c}{TMD60(O)}\\
\hline\hline
\               & \ \ MD\ \ & \ MAD\ & \ RMSD & \ minD  & \ maxD \\ [0.5ex]
\hline
OS-CCSD-SPT(2)  &  -1.92    &  1.92  &  2.27  &  -4.14  & -0.41  \\
OS-CCSD-SPT(3)  &   1.82    &  3.31  &  4.14  &  -6.80  &  7.46  \\
CCSD            & -15.06    & 15.06  & 15.83  & -19.91  & -5.00  \\
CCSD(T)         &   3.03    &  3.03  &  3.71  &   0.20  &  8.30  \\
[1ex]
\hline\noalign{\medskip}
\multicolumn{6}{c}{TMD60(S)}\\
\hline\hline
\               & \ \ MD\ \ & \ MAD\ & \ RMSD & \ minD  & \ maxD \\ [0.5ex]
\hline
OS-CCSD-SPT(2)  &  -1.10    &  1.27  &  1.58  &  -2.87  &  0.47   \\
OS-CCSD-SPT(3)  &  -0.84    &  2.26  &  3.20  &  -6.76  &  2.83   \\
CCSD            & -10.53    & 10.53  & 11.02  & -14.68  & -5.33   \\
CCSD(T)         &   2.23    &  2.23  &  2.52  &   1.03  &  4.11   \\
[1ex]
\hline
\end{tabular}
\label{tableTMD60}
\end{table}

The use of PBE0 orbitals leaves some residual spin contamination, e.g.\ $\langle S^2 \rangle = 0.86$ for ZnO (singlet), and 
$\langle S^2 \rangle = 6.56$ for CrS (quintet).  Thus, it is not surprising that the deviations of OS-CCSD-SPT(3)
and CCSD(T) from CCSD(2) are larger than for W4-08woMR.  Nevertheless, OS-CCSD-SPT(2) exhibits the smallest deviations
from CCSD(2) for these datasets.  Despite the significance of this result, we must remark that although CCSD(2) is the best of the above
models for these TMD60 subsets,\cite{HaitJCTC2019} it does not provide an overall convincing benchmark here, and we caution against the idea of 
any of these approaches being suitable for MR situations.

\begin{figure}[!b]
\begin{center}
\includegraphics[scale=0.5]{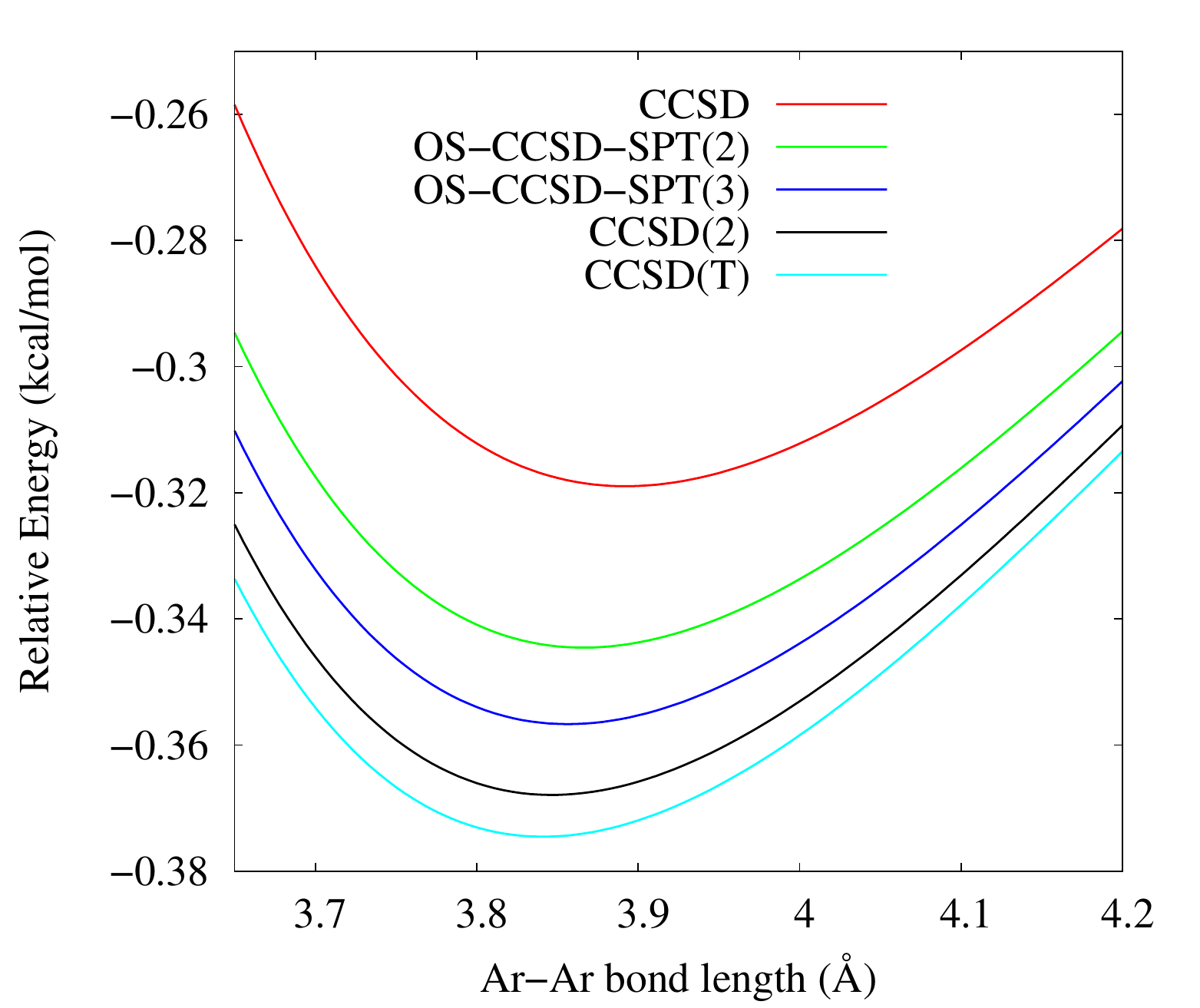}
\caption[]{Ar$_2$ PES.  All calculations used HF orbitals and CCSD(2) used $\bar{F}$ for $H_0$.}
\label{ar2plot}
\end{center}
\end{figure}

\begin{figure}[!b]
\begin{center}
\includegraphics[scale=0.5]{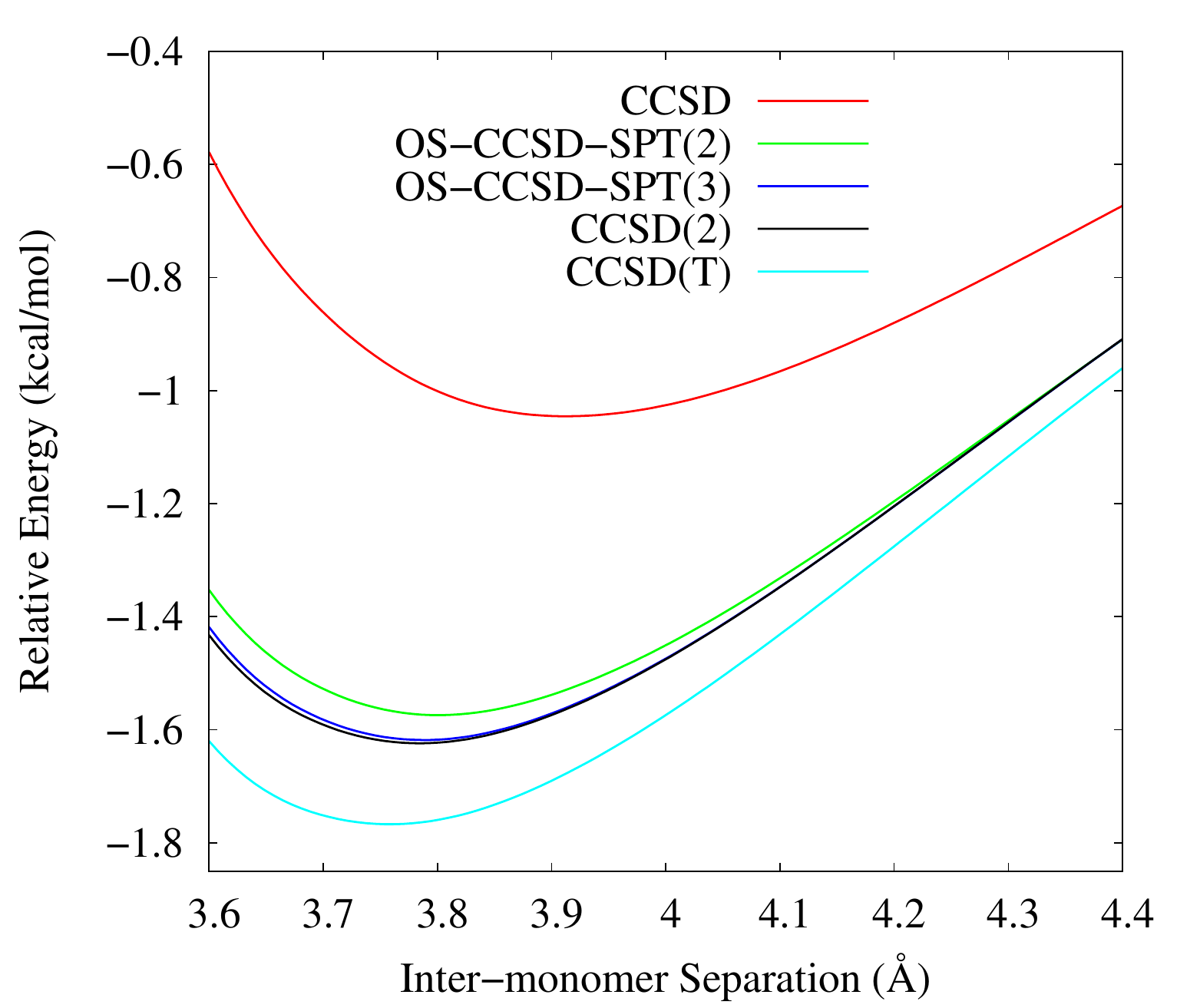}
\caption[]{Sandwich benzene dimer PES.  All calculations used HF orbitals (def2-SV(P)) and CCSD(2) used $\bar{F}$ for $H_0$.}
\label{bz2plot}
\end{center}
\end{figure}

Finally, to test the effectiveness of OS-CCSD-SPT(2) for non-covalent interactions, we did calculations on the argon dimer and sandwich benzene dimer.  
The latter is the parallel stacking of the two monomers, and we examined several inter-monomer distances, using the same geometries considered in 
ref.\ \onlinecite{TakataniJCP2008}.  The results for these two PESs are given in Figs.\ \ref{ar2plot} and \ref{bz2plot}.

For the benzene dimer,
OS-CCSD-SPT(2) is close to CCSD(2) (and OS-CCSD-SPT(3) is nearly indistinguishable from CCSD(2) here), while OS-CCSD-SPT(2)'s improvement over CCSD is
less dramatic, although still substantial, for Ar$_2$.  In any case, it is clear that the benefits of OS-CCSD-SPT(2) extend to the non-covalent realm.
In passing, and because it is a larger molecule than any in the above 3 datasets, we show the AEs (kcal/mol) for C$_6$H$_6$:
OS-CCSD-SPT(2): 1282.59, OS-CCSD-SPT(3): 1283.36, CCSD: 1267.47, and CCSD(2): 1283.46.

To summarize this work, we have presented OS-CCSD-SPT(2), a very simple modification to the standard CC approach that is free of empirical parameters and 
whose computational cost is overall the same as that of CCSD (but with a lower prefactor), yet is able to produce an accuracy level that is 
competitive with that of much more complicated models like CCSD(T). 

Future work should include efforts to better understand the reasons for OS-CCSD-SPT(2)'s success.  This might entail investigating generalizations to higher
order CC methods like CCSDT, where the goal might be to circumvent perturbative corrections like CCSDT(Q)\cite{KucharskiJCP1998_2,BombleJCP2005} 
or CCSDT(Q)$_{\Lambda}$.\cite{KallayJCP2005}
Or it might entail smaller adjustments to the structure of the present approach.  For this,
we thought it worthwhile to assess ``singlet-paired'' CCSD, i.e.\ CCSD0,\cite{BulikJCTC2015} in the context of the present work.
The motivation for this is clear: CCSD0 can be viewed as a spin-symmetric reduction of OS-CCSD.
It was therefore surprising that applying SPT to CCSD0 appears to lead to worse performance: for example, the binding energy for
the above sandwich benzene dimer is around 2.05 kcal/mol at $E^{(2)}$ and 1.38 kcal/mol at $E^{(3)}$, with CCSD(2) lying in between these values.
There are other possibilities for fine-tuning OS-CCSD-SPT(2), e.g.\ 
orbital optimization,\cite{SherrillJCP1998,NeeseJCTC2009,BozkayaJCP2011}
spin-component scaling,\cite{GrimmeWCMS2012} and 
regularization.\cite{LeeJCTC2018,BertelsJPCL2019}  

Also for future work,
more benchmark studies, including energies, geometries, vibrational frequencies, etc., will be necessary to further pinpoint OS-CCSD-SPT(2)'s efficacy.
We are particularly interested in applying the various linear-scaling approximations mentioned above to OS-CCSD-SPT(2).  Because OS-CCSD is a simple
sub-model of CCSD, this should be quite straightforward.  This could extend the applicability of accurate CC calculations to an even larger range, thus
e.g.\ reducing the disparity between CC and DFT, and expanding the use of CC as a foundation for training 
Machine Learning methods.\cite{NudejimaJCP2019,ChengJCP2019}

Earlier works have demonstrated higher than expected accuracy in triples-free approaches.\cite{HuntingtonJCP2012,SedlakCPC2013,BertelsJPCL2019}
Similarly, the OS-CCSD-SPT(2) results shown in this Letter along with the method's prospects for future development
strongly help to contend that perturbative triples 
(and quadruples) may not be necessary to achieve chemical accuracy in coupled cluster calculations.

\section{acknowledgement}
The author thanks Dr. Kathleen Durkin and Xiaowei Xie for encouragement.  Calculations were performed at the UC Berkeley Molecular Graphics and Computation Facility (MGCF).  MGCF is supported by grant NIH S10OD023532.

\bibliography{ms.bbl}

\end{document}